\documentclass[aps,superscriptaddress,prl,twocolumn,floatfix]{revtex4}
\usepackage{multirow}
\usepackage{graphicx}
\usepackage{bbm}
\usepackage{amsmath,amssymb,array}
\usepackage[usenames]{color}

\newcommand{\bra}[1] { \langle #1 | }
\newcommand{\ket}[1] { | #1 \rangle }
\renewcommand{\vr} {{\bf r}}

\def\G{\hat{\Gamma}}
\def\Gs{\hat{\Gamma}\s}

\def\F{F}
\def\ees{_{\sss ee,s}}
\def\g{_\gamma}

\def\t{^\tau}
\def\LDAt{^{{\rm LDA}\tau}}
\def\GGAt{^{{\rm GGA}\tau}}
\def\unift{^{{\rm unif}\tau}}

\def\tl{^{\tau,\lambda}}
\def\tp{^{\tau'}}
\def\tg{^{\{\tau/\gamma^2\}}}

\def\tg{^{\tau/\gamma^2}}

\def\n{n}
\def\Eq#1{Eq. (\ref{#1})}

\def\bea{\begin{eqnarray}}
\def\eea{\end{eqnarray}}
\def\ben{\begin{equation}}
\def\een{\end{equation}}
\def\benu{\begin{enumerate}}
\def\enu{\end{enumerate}}

\def\bei{\begin{itemize}}
\def\eei{\end{itemize}}
\def\beit{\begin{itemize}}
\def\eit{\end{itemize}}
\def\benu{\begin{enumerate}}
\def\enu{\end{enumerate}}

\def\n{n}

\def\sss{\scriptscriptstyle\rm}

\def\g{_\gamma}

\def\1var{(\bx_1...\bx\N)}

\def\br{{\bf r}}

\def\bx{{x}}

\def\x{_{\sss X}}
\def\c{_{\sss C}}
\def\s{_{\sss S}}
\def\xc{_{\sss XC}}

\def\N{_{\sss N}}

\def\unif{^{\rm unif}}

\def\ee{_{\rm ee}}

\def\sph_int{ {\int d^3 r}}

\begin{document}
\title{Exact conditions in finite temperature density functional theory}
\author{S. Pittalis}
\email[Electronic address:\;]{pittaliss@missouri.edu}
\affiliation{Department of Physics and Astronomy, University of Missouri-Columbia,
Columbia, Missouri 65211}
\affiliation{Institut f{\"u}r Theoretische Physik,
 Freie Universit{\"a}t Berlin, Arnimallee 14, D-14195 Berlin, Germany}
\affiliation{European Theoretical Spectroscopy Facility (ETSF)}
\author{C. R. Proetto}
\altaffiliation[Present address: ]{Centro At{\'o}mico Bariloche and Instituto Balseiro, 8400
S.C. de Bariloche, R{\'i}o Negro, Argentina.}
\affiliation{Institut f{\"u}r Theoretische Physik,
Freie Universit{\"a}t Berlin, Arnimallee 14, D-14195 Berlin, Germany}
\affiliation{European Theoretical Spectroscopy Facility (ETSF)}
\affiliation{Max-Planck-Institut f{\"u}r Mikrostrukturphysik, 
Weinberg 2, D-06120 Halle, Germany}
\author{A. Floris}
\altaffiliation[Present address:]{ Department of Physics, King's College 
London, London, Strand WC2R 2LS United Kingdom.}
\affiliation{Institut f{\"u}r Theoretische Physik,
Freie Universit{\"a}t Berlin, Arnimallee 14, D-14195 Berlin, Germany}
\affiliation{European Theoretical Spectroscopy Facility (ETSF)}
\author{A. Sanna}
\affiliation{European Theoretical Spectroscopy Facility (ETSF)}
\affiliation{Max-Planck-Institut f{\"u}r Mikrostrukturphysik, 
Weinberg 2, D-06120 Halle, Germany}
\author{C. Bersier}
\affiliation{European Theoretical Spectroscopy Facility (ETSF)}
\affiliation{Max-Planck-Institut f{\"u}r Mikrostrukturphysik, 
Weinberg 2, D-06120 Halle, Germany}
\author{K. Burke}
\affiliation{Department of Chemistry, UC Irvine, CA 92697}
\author{E. K. U. Gross}
\affiliation{European Theoretical Spectroscopy Facility (ETSF)}
\affiliation{Max-Planck-Institut f{\"u}r Mikrostrukturphysik, 
Weinberg 2, D-06120 Halle, Germany}
\date{\today}

\begin{abstract}
Density functional theory (DFT) for electrons at finite temperature is increasingly
important in condensed matter and chemistry. 
Exact conditions that have proven crucial
in constraining and constructing accurate approximations for ground-state DFT are
generalized to finite temperature, including the adiabatic connection formula.
We discuss consequences for functional
construction.
 \end{abstract}


\maketitle

Because of the small mass ratio between electrons and nuclei,
standard electronic structure calculations treat the former
as being in their ground state, but routinely account for
the finite temperature of the latter, as in {\em ab initio} molecular
dynamics~\cite{CP85}.  
But as electronic structure methods are applied  in ever more esoteric areas, the need
to account for the finite temperature of electrons increases.
Phenomena where such effects play a role include
rapid heating of solids via strong laser fields~\cite{GRTS09},
dynamo effects in giant planets~\cite{RMNF11},
magnetic~\cite{A,B} and superconducting phase transitions~\cite{C,D}, shock
waves~\cite{RMCH10,M06}, warm dense matter~\cite{KRDM08}, and hot plasmas~\cite{DP82,DM08,PD00}. 

Within density functional theory, the natural framework for
treating such effects was created by Mermin~\cite{M65}.
Application of that work to the Kohn-Sham (KS) scheme
at finite temperature also yields a natural approximation:
treat KS electrons at finite temperature but use ground-state 
exchange-correlation (XC) functionals.  This works
well in recent calculations~\cite{KRDM08,RMCH10}, where inclusion
of such effects is crucial for accurate prediction.
This assumes that finite-temperature effects on
exchange-correlation 
are negligible relative to the KS contributions,
which may not always be true.

The uniform electron gas at finite temperature (also called
the one-component plasma) has been well-studied, and has in the
past provided the natural starting point for DFT studies of such
finite-temperature XC effects, as input into the local density
approximation (LDA) at finite $T$ \cite{DAC86}.  
However,  the LDA is too
inaccurate for most modern applications of DFT, and almost
all recent calculations use a generalized gradient approximation
or hybrid with exchange \cite{FNM03}.   The errors of LDA would typically
be enormous relative to the temperature corrections we seek,
especially for correlation, and so could lead to quite misleading
results.  Accurate calculation of
finite temperature contributions requires accurate approximate
functionals.
Magnetic phase transitions bear an additional difficulty:
The low-lying excitations are collective, i.e., magnons whose
description requires non-collinear
version of spin-DFT. Hence, a finite-temperature version
of spin-DFT involving only spin-up and spin-down
densities and thus only spin-flip excitations, is bound to fail in predicting,
e.g., the critical temperature~\cite{A}.

The most fundamental steps toward both understanding
a functional and creating accurate approximations are deriving its
inequalities from the variational definition of the functional.
These yield both the signs of energy contributions and, via
uniform scaling of the spatial coordinates, basic equalities and
inequalities that non-empirical functionals should satisfy by construction.
The adiabatic connection formula~\cite{LP79} is intimately related.
Here, we (i) establish the fundamental functionals needed
for treating finite temperature, (ii) prove the most basic properties
(signs of the energy contributions), (iii) show that the temperature
must be scaled simultaneously with the spatial coordinate, (iv)
derive the inequalities under such scaling, and (v) give the adiabatic
connection formula for finite temperature.  These results establish
the basic rules for all finite-temperature KS treatments.

Central to the thermodynamic description of many-electron systems
is the grand-canonical potential, defined as the statistical 
average of the grand-canonical operator
\begin{equation}\label{gcop}
\hat{\Omega} = \hat{H}  - \tau \hat{S} - \mu \hat{N},
\end{equation}
where $\hat{H}$, $\hat{S}$, $\hat{N}$, $\tau$ and $\mu$ are the Hamiltonian, entropy, and particle-number operators,
temperature and chemical potential, respectively. In detail, $\hat{H} =  \hat{T} + \hat{V}_{\rm ee} + \hat{V}$,   
where $\hat{T}$ and  $\hat{V}_{\rm ee}$  are the kinetic energy and the
Coulomb electron-electron interaction operators, and $\hat{V}$ represents 
an external scalar potential $v(\vr)$. The entropy operator is given by
$\hat{S} = - ~ k \ln \hat{\Gamma} \; ,$
where $k$ is the Boltzmann constant and $\hat{\Gamma}=  \sum_{N,i} {w_{N,i}} \ket{\Psi_{N,i}} \bra{\Psi_{N,i}}$ 
is a statistical operator, with $\ket{\Psi_{N,i}}$ and $w_{N,i}$ being orthonormal 
$N$-particle states and statistical weights, respectively, with the latter satisfying the (normalization) condition 
$\sum_{N,i} w_{N,i} = 1$.
The statistical average of an operator $\hat{A}$ is obtained as
\begin{equation}
A[\hat{\Gamma}] =  {\mbox{Tr}} ~ \{\hat{\Gamma}\hat{A}\} = \sum_N \sum_i w_{N,i} \bra{\Psi_{N,i}} \hat{A} \ket{\Psi_{N,i}} \; . 
\end{equation}

The thermodynamical equilibrium properties of many-electron systems are
obtained from the knowledge of the grand-canonical statistical operator 
$
\hat{\Gamma}^0 = \sum_{N,i} {w}_{N,i}^0
\ket{\Psi_{N,i}^0} \bra{\Psi_{N,i}^0},
$
where $\ket{\Psi_{N,i}^0}$ are the $N$-particle eigenstates of $\hat{H}$ with energies ${E}_{N,i}^0$,
and the equilibrium statistical weights are given by  
$w_{N,i}^0 = \tfrac{ \exp[-\beta({E}_{N,i}^0-\mu N)] }{\sum_{N,i}  \exp[-\beta({E}_{N,i}^0-\mu N)] }$, 
where $\beta=\tfrac{1}{k\tau}$~\cite{PY89}. 
The Gibbs principle ensures that $\hat{\Gamma}^0$ minimizes the statistical average of the grand-potential operator.
We emphasize that $\hat{\Gamma}^0$ is unique~\cite{M65} and that
in the limit of zero temperature, for systems with degenerate ground states,
it leads to ensembles with equal statistical weights.

To create a DFT at finite temperature, Mermin~\cite{M65} rewrites this as
(in modern parlance)
\ben
\Omega\t_{v-\mu} = \min_\n \left\{\F^\tau[\n] + \int d^3r\, \n(\br)\, (v(\br)-\mu)
\right\}
\label{Mermin}
\een
where the minimizing $n(\br)$ is the equilibrium density $n^0(\br)$, and 
\ben
\F\t[\n] :=  \min_{\G\to\n}  \F\t [\G]  = \min_{\G\to\n} \left\{ T[\G] + V\ee[\G] - \tau S[\G]\right\},
\label{Ft}
\een
is the finite-temperature analog of the universal Hohenberg-Kohn functional,
defined through a constrained search~\cite{L79,PY89}. 
This depends only on $\tau$ and not on $\mu$.
We denote $\G\t[\n]$ as the minimizing statistical operator in Eq.~(\ref{Ft}), and define the
density functionals:
\bea
T\t[\n]&:=&T[\G\t[\n]],~~~
V\ee\t[\n]:=V\ee[\G\t[\n]],\nonumber\\
S\t[\n]&:=&S[\G\t[\n]],
\eea
i.e., each density functional is the trace of its operator
over the minimizing $\G$ for the given $\tau$ and density. 

Next consider a system of non-interacting
electrons at the same 
temperature $\tau$, and denote its one-body potential as
$v\s(\br)$.  All the previous arguments apply, and we choose $v\s(\br)$ 
to make its density match that of the interacting problem.  This defines the
KS system at finite temperature.
Because it arises so often
in this work, we define the kentropy as
\ben
K\t[\G] := T[\G] - \tau\, S[\G],
\label{kentropy}
\een
and we show it plays an analogous role to the kinetic energy in ground-state
DFT, to which it reduces as $\tau\to 0$.
The non-interacting functional is just
\ben
F\s\t[\n] :=  \min_{\G\to\n} K\t[\G] = K\t[\Gs\t[\n]]
\label{Fst}
\een
from \Eq{Ft} applied without $V\ee$,
and we define:
\ben
T\s\t[\n]:=T[\Gs\t[\n]] \; ,  ~~~
S\s\t[\n]:=S[\Gs\t[\n]].
\een 
Next we define the difference functionals that are crucial to the KS method.  
Write
\ben
V\ees\t[\n]:=V\ee[\Gs\t[\n]]=U\t[\n]+\Omega\x\t[\n] \; ,
\een
where $U\t[\n]$ in terms of the density has the form of the usual Hartree energy, 
and expressing $\Omega\x\t[\n]$ in terms of the module square of the 
one-body density matrix stemming from $\Gs\t[\n]$~\cite{GCG10} we observe that $\Omega\x\t[\n] \leq 0$.

The kinetic correlation is
\ben
T\c\t[\n] :=  T[\G\t[n]]-T[\Gs\t[\n]] \; ,
\een
and similarly define
$ S\c\t[\n]$ and 
$ K\c\t[\n]$,
while the potential contribution is
\ben
U\c\t[\n] := V\ee[\G\t[\n]] - V\ee[\Gs\t[\n]].
\een
The sum of the energy components is, as in ground-state DFT, the correlation energy,
$E\c\t[\n] := T\c\t[\n] + U\c\t [\n]$, 
while the grand-canonical correlation potential is
\ben
\Omega\c\t[\n] := K\c\t[\n] +U\c\t[\n] = E\c\t[\n] - \tau S\c\t[n] \; ,
\label{Oct}
\een
and
$\Omega\xc\t[\n] := \Omega\x\t[\n] + \Omega\c\t[\n]$.

We now prove the most basic theorems about the signs of our quantities.
To show that the correlation-kentropy (or kentropic correlation) is
always positive, begin by noting 
$K\t[\Gs\t[\n]] \leq K\t[\G\t[\n]]$, because 
$\G\s\t[\n]$ minimizes $K\t[\G]$.
By inserting the definition, \Eq{kentropy}, we find
$K\c\t[\n] \geq 0$,
with equality only when the interaction is zero.  
It is the kentropic correlation that is guaranteed
to be positive, not the kinetic correlation alone, contrary
to the pure ground-state case\cite{LP85}.
Similarly, since
$F\t[\G\t[\n]] \leq F\t[\Gs\t[\n]]$,
we find
$\Omega\c\t[\n] \leq 0$.
Combining these results with \Eq{Oct} implies
$U\c\t[\n] \leq 0$.
Thus
\ben
\Omega\x\t[\n] \leq 0,\; \Omega\c\t[\n] \leq 0,\; U\c\t[\n] \leq 0,\ K\c\t[\n] \geq 0,
\label{sum}
\een
and no approximation should violate these basic rules.

Some of the most important results in ground-state DFT
come from uniform scaling of the coordinates\cite{LP85,PBE96}.  
In the following considerations, when we refer explicitly to wavefunctions, we shall restrict to
wavefunctions having finite norm on their entire domain of  definition.
Under norm-preserving homogeneous scaling of the coordinate $\vr \rightarrow \gamma \vr$, with
$\gamma>0$, to the scaled wave function~\cite{LP85}
\begin{equation}\label{scwav}
\Psi^{\gamma}(\vr_1,...,\vr_N) := \gamma^{\frac{3}{2}N} \Psi(\gamma\vr_1,...,\gamma\vr_N),
\end{equation}
corresponds the scaled density $n_{\gamma}(\vr)=\gamma^3 n(\gamma \vr)$.
Writing $\Psi^{\gamma}(\vr_1,...,\vr_N) = \langle \vr_1,...,\vr_N | \Psi^{\gamma} \rangle$
in terms of the (representation-free) element $\ket{\Psi^{\gamma}}$ of Hilbert space, the scaled 
statistical operator is defined as
\begin{equation} \label{scaling}
\hat{\Gamma}_{\gamma} := \sum_{N} \sum_{i} w_{N,i} \ket{\Psi_{N,i}^{\gamma}} \bra{\Psi_{N,i}^{\gamma}}\;,
\end{equation}
where the statistical weights are hold fixed, 
i.e., the scaling only acts on the states.   

With the above definition, the statistical
average of an operator whose pure-state expectation value
scales homogeneously \cite{LP85}, scales homogeneously as well. 
In particular, we have: $T[\hat{\Gamma}_{\gamma}] = \gamma^2 T[\hat{\Gamma}]$, 
$V_{\rm ee}[\hat{\Gamma}_{\gamma}] = \gamma V_{\rm ee}[\hat{\Gamma}]$, 
$N[\hat{\Gamma}_{\gamma}] = N[\hat{\Gamma}]$, and $S[\hat{\Gamma}_{\gamma}]=S[\hat{\Gamma}]$.
The scaling behavior of the {\it density} functionals is, however, more subtle.
First consider  the non-interacting functionals in some detail.
Because $\Gs\t[n]$ minimizes $K\t$, \Eq{Fst}, and
\ben
K\t[\G\g] = \gamma^2 \left( T[\G] - \frac{\tau}{\gamma^2}\, S[\G] \right) = \gamma^2 K^{\tau/\gamma^2}[\G] \; ,
\label{KtGg}
\een
then
\ben
\Gs\t[\n\g]= \hat{\Gamma}^{\tau/\gamma^2}_{{\s},\gamma} [\n] \; ,~~~~
F\s\t[\n\g]= \gamma^2\, F\s\tg[\n] \; .
\label{Fstng}
\een
In particular we notice that
\ben
S_s^{\tau}[\n\g] = S^{\tau / \gamma^2}_s[\n] \; .
\label{Sstg}
\een
For non-interacting electrons, the statistical operator at
a given temperature that is the minimizer for a given {\em scaled}
density is simply the scaled statistical operator, but at a {\em scaled}
temperature, an effect that is obviously absent in the ground-state
theory.

There are further simple implications.  First, if we invert the
sense of \Eq{Fstng}, we can write:
\ben
F\s\tp[\n]= \frac{\tau'}{\tau}\, F\s\t[\n_{{\sqrt{\tau/\tau'}}}],
\een
i.e., knowledge of $F\s\t[\n]$ at any one finite $\tau$ generates
{\em its entire temperature dependence}, via scaling.
Furthermore, it must always collapse to the
ground-state KS kinetic energy under scaling to the high-density limit:
\ben
T\s[\n]  = \lim_{\gamma\to\infty} F\s\t[\n\g]/\gamma^2.
\een
Similarly, in the low-density limit
\ben
S\s^\infty[\n]
= -
\lim_{\gamma\to 0 } F\s\t[\n\g]/\tau,
\een
where $S\s^\infty[\n]$ is the non-interacting KS entropy in the high-temperature limit.

Next, we consider the interacting case.  The exchange contribution is
much simpler than correlation, because it is extracted from the
one-particle density matrix.  
Because $V_{\ee}[\G]$ and $U[\G]$ 
scale
linearly with $\gamma$, and using the simple scaling relation for $\Gs$, \Eq{Fstng},
\ben
\Omega\x\t[\n\g] = \gamma\, \Omega\x\tg[\n] \; .
\label{Exng}
\een
This scaling result is important in ground-state DFT, where it restricts the dependence of
the exchange-enhancement factor to depending on just the reduced density
gradient~\cite{PBE96}.
But the more interesting case is correlation.
From the definition, Eq. (\ref{Ft}),
\ben
F\t[\n\g] \leq F\t[\G\g\tp[\n]] \; ,
\een
since $\G\g\tp[\n]$ has density $\n\g$, and $\tau'$ is {\em any} temperature.
Using the scaling properties and choosing $\tau'=\tau/\gamma^2$, then the fundamental
inequality of scaling is
\ben
K\t[\n\g]+V\ee\t[\n\g] \leq 
\gamma^2\, K\tg[\n]+\gamma\, V\ee\tg[\n]  .
\label{Ktineq}
\een
To find a condition on the kentropy alone, define $\n'(\br)=\n\g(\br)$,
$\gamma'=1/\gamma$, and $\tau'=\tau/\gamma^2$ in \Eq{Ktineq}.
Multiply the result by $\gamma'$, and combine with \Eq{Ktineq}, to find
\ben\label{Kineq}
K\t[\n\g] \leq \gamma^2\, K\tg[\n],~~~~~\gamma \ge 1 \; .
\een
This is the finite temperature
analog of the subquadratic scaling of the kinetic energy in the real system~\cite{LP85}.
Another combination isolates the repulsive contributions:
\ben\label{Veeineq}
V\ee\t[\n\g] \geq \gamma\, V\ee\tg[\n],~~~~~\gamma \ge 1 \; .
\een
These inequalities loosely constrain the behavior of these large
energies.  Much more important is to subtract out KS quantities that
scale simply, to find for $\gamma \ge 1$:
\ben\label{Kcineq}
K\c\t[\n\g] \leq \gamma^2\, K\c\tg[\n],~~~~~
U\c\t[\n\g] \geq \gamma\, U\c\tg[\n].
\een
One more application of \Eq{Ktineq} yields
\ben
\Omega\c\t[\n\g] \geq \gamma\, \Omega\c\tg[\n],~~~~~\gamma \ge 1,
\label{Ocineq}
\een
the fundamental scaling inequality for the correlation
contribution to the grand canonical potential.
The inequalities, Eqs. (\ref{Kineq}-\ref{Ocineq}), which
are reversed if $\gamma < 1$, 
provide tight constraints on these
functionals and are routinely used in non-empirical functional construction in
the ground state\cite{PBE96}.
For example, combining \Eq{Exng} with \Eq{Ocineq} in the high-density limit, yields:
\ben
\Omega\t\x[\n]=\lim_{\gamma\to\infty} 
\Omega\xc^{\gamma^2\tau}[\n\g]/\gamma.
\label{Exlimdef}
\een
This scaling procedure can usually be applied easily
to any approximate $\Omega\xc\t[\n]$ to extract its separate
exchange and correlation contributions.

Lastly, we consider the adiabatic coupling constant for finite temperature,
its relationship to scaling, 
and derive the adiabatic connection formula. 
Define
\ben
\F\tl[\n] = \min_{\G\to\n} \left\{ T[\G] +\lambda\, V\ee[\G] - \tau S[\G]\right\} \; ,
\label{Ftlambda}
\een
with $ \hat{\Gamma}^{\tau,\lambda}[n]$ being the corresponding minimizing $\hat{\Gamma}$.
By scaling, it is straightforward to show:
\ben
\G\tl[\n] = \G^{\tau / \lambda^2}_\lambda [\n_{1/\lambda}] \; ,~~~~~
F\tl[\n] = \lambda^2\, F^{\tau / \lambda^2}[\n_{1/\lambda}] \; .
\label{Ftl}
\een
where quantities with one superscript are evaluated at $\lambda=1$.
\Eq{Ftl} is the interacting generalization of \Eq{Fstng} and
shows that, even in the presence of interactions, simple equalities
are possible, but at the price of altering the coupling constant. In particular we notice that
\ben
S\tl[\n] = S^{\tau / \lambda^2}[\n_{1/\lambda}] \; .
\label{Stl}
\een
Of course, non-interacting functionals are not affected by a coupling constant modification. 
\Eq{Exng} implies that the exchange and Hartree density functionals have a linear dependence on $\lambda$.
Employing the minimization property of \Eq{Ftlambda} and the Hellmann-Feynman theorem,
we find
\ben
\Omega\xc\t[\n] = \int_0^1 d
\lambda\, U\xc\t[n](\lambda) \; ,
\label{ACF}
\een
where
\ben
U\xc\t[n](\lambda)=V\ee[\G\tl[\n]]-U\t[\n] \;.
\een
Eq.~(\ref{ACF}) is the finite-temperature adiabatic connection formula, whose 
zero-temperature limit played a central role in ground-state DFT.
$U\xc\t[n](0) = \Omega\x\t[n] < 0$ (Eq.(13)), and the scaling inequalities can be 
combined, analogously to Ref.~\cite{LP85}, to show 
that $U\xc\t[n](\lambda)$ is monotonically decreasing in $\lambda$.

So far, all results presented have been exact.  To see them 
in practice, consider the finite-temperature local density approximation (LDA) to $ \Omega\xc\t[n]$
\ben
\Omega\xc\LDAt[\n] = \int d^3r\, \omega\xc\unift(\n(\br)) \; ,
\een
where $\omega\xc\unift(n)$ is the XC grand canonical potential density of
a uniform electron gas of density $\n$.  
Because a {\it uniform} electron gas is a quantum mechanical system, its energies
satisfy all our conditions, guaranteeing 
by construction that LDA satisfies all the
exact conditions listed here.  In the Jacob's ladder of functional construction~\cite{FNM03}, 
more sophisticated approximations should also
satisfy these conditions.
To give one simple example, \Eq{Exng} implies
\ben
\omega\x\unift(\n(\vr)) = e\x\unif(n(\vr)) \, F\x(\tilde\tau(\vr)) \; ,
\een
where $e\x\unif(n(\vr)) = - A\x \; n^{4/3}(\vr) $, $A\x = (3/4 \pi)(3 \pi^2)^{1/3}$, and
$\tilde\tau(\vr)=\tau/\n^{2/3}(\vr)$ is a dimensionless measure of the local
temperature.  Thus the largest fractional deviations from ground-state results
should occur (in LDA) in regions of lowest density, but these contribute less
in absolute terms.
For a generalized gradient approximation (GGA), \Eq{Exng} implies 
\ben
\omega\x\GGAt(\n(\vr),|\nabla n|(\vr)) = e\x\unif(\n(\vr))\, F\x(s(\vr),\tilde\tau(\vr)),
\een
where the dimensionless gradient $s$ is $|\nabla n|/(2 k_F \n)$
and $k_F=(3\pi^2 \n)^{1/3}$,
i.e., the exchange enhancement factor $F\x(s,\tilde\tau)$
depends on the temperature only via $\tilde\tau$.

In summary, 
there is a present lack of approximate density functionals for finite temperature.
We have derived many
basic relations needed to construct such approximations, and expect future
approximations to either build these in, or be tested against them.
In principle, such approximations should already be implemented in
high-temperature DFT calculations, at least at the LDA level, as a check
that XC corrections due to finite temperature do not alter calculated results.
If they do, then more accurate approximations than LDA will be needed to
account for them.

This work was supported by the Deutsche Forschungsgemeinschaft. C.R.P. was supported by the European Community 
through a Marie Curie IIF (Grant No. MIF1-CT-2006-040222). S.P. acknowledges 
support through DOE grant DE-FG02-05ER46203. 
K.B. was supported through DOE grant DE-FG02-08ER46496.

\label{page:end}
\bibliography{T}
\end{document}